\documentclass[useAMS,usenatbib,usegraphicx]{mn2e}

\title[The spectroscopic variability of T$_{\rm eff}$ and $\log g$ in $\beta$~Cep]
  {The spectroscopic variability of T$_{\rm eff}$ and $\log g$ in
        the hot pulsating star $\beta$ Cephei}
\author[G. Catanzaro, F. Leone]{G.~Catanzaro${^1}$\thanks{E-mail: Giovanni.Catanzaro@oact.inaf.it}
        and  F.~Leone${^2}$\\
  ${^1}$INAF - Osservatorio Astrofisico di Catania, Via S. Sofia 78, 95123 Catania, Italy\\
  ${^2}$Universit\`a di Catania, Dipartimento di Fisica e Astronomia - Sezione Astrofisica,  
   Via S. Sofia 78, 95123 Catania, Italy}

\begin{document}

\date{Accepted 2008 July 2;  Received 2008 July 2; in original form 2008 April 2}
\pagerange{\pageref{firstpage}--\pageref{lastpage}} \pubyear{2008}

\def\LaTeX{L\kern-.36em\raise.3ex\hbox{a}\kern-.15em
    T\kern-.1667em\lower.7ex\hbox{E}\kern-.125emX}

\label{firstpage}

\maketitle

\begin{abstract}
Time resolved spectra of $\beta$\,Cep yield the following average results:
T$_{\rm eff}$\,=24000\,$\pm$\,250~K, $\log g$\,=\,3.91\,$\pm$\,0.05 and 
$\xi$\,=\,8.1\,$\pm$\,0.9~km~s$^{-1}$. N, O, Ne, Al, Si and S abundances are 
solar while C, Mg and Fe are slightly under-abundant.
Pulsational amplitudes of $\Delta$T$_{\rm eff}$\,$\sim$\,700~K and
$\Delta$$\log g$\,$\sim$\,0.2 dex are found from H$\beta$. The metal
lines give similar amplitudes but centred on T$_{\rm eff}$\,$\sim$\,25000~K.
An upper limit of 1.0 km s$^{-1}$ to the variability of the microturbulence 
is derived from the Si {\sc iii} triplet at 455~nm. The radial velocity
amplitude derived from the core of H$\beta$ is $\sim$\,15$\%$ greater
than that from the metal lines.

\end{abstract}


\section{Introduction}
The capability of oscillations to probe the internal structure of stars
is fully testified by solar studies. For this reason, 
pulsating stars, that are spread in the HR diagram, are expected to
largely improve our knowledge of stellar evolution and physical assumptions in
models. 

\citet{osaki71} has shown that both radial and non-radial modes of
pulsating stars  are observationally distinguishable via the analysis of line profile
variation (LPV). From this pioneering work, there has been an increasing
interest in LPV modeling driven by the improvement in resolution and
efficiency of modern spectrographs: \citet{lee90} included the
influence of stellar rotation, \citet{briquet01} considered
photospheric spots, \citet{aerts94} performed a mode identification via the moments 
of the line profile or \citet{telting97} via the Doppler imaging technique. 

\citet{dupret02} noted that a pure kinematic approach is not
enough for LPV modeling and that  
a realistic representation of the strength and broadening of spectral lines
were necessary. They considered the atmosphere of a pulsating
star as a perturbed status of the hydrostatic equilibrium and represented
the temperature perturbation of atmospheric layers with a series expansion
in terms of effective temperature and gravity. 

This operative method does not clearly represent the physics of $\beta$\,Cep
stars, pulsation modes do not change the energy production and the very small
(1\%) stellar radius variation \citep{aerts94} cannot modify
neither the effective temperature nor the surface gravity in an observable way.  
However along a pulsation cycle, the compression and expansion of photospheric
layers
change the dependence of temperature and pressure on optical depth and let 
the effective temperature appear variable, with the pulsation period,
if determined in a limited spectral interval. A phenomenon indeed observed
in some $\beta$ Cephei stars by \citet{deridder02} on the basis of a photometric 
approach. These authors derived a T$_{\rm eff}$ variability up to 1000 K with 
the pulsation period, from Geneva colours.

With the density and temperature, the variation of atomic level
populations should also result in apparent surface gravity variations.
In fact, there is an equivalence between stellar surface gravity and
electron pressure in the framework of spectral line formation \citep{Gray92}.

A further contribution to the strength and profile of spectral lines comes
from the microturbulence ($\xi$), that in principle could be modified along 
compression and expansion phases.

In this paper we analyse a set of time resolved spectra of $\beta$ Cep 
(=~HD\,205021~=~HR\,8238), which is generally accepted to be the proto-type of
a class of hot pulsating stars, with the aim of ascertaining the apparent
variability of the effective temperature, gravity and
microturbulence with the pulsation period. These variations would
strongly constrain any LPV method, as the one proposed by Dupret and collaborators.

\begin{table}
\caption{For each spectrum we report the Heliocentric Julian Date,
the radial velocity measured from the Si {\sc iii} triplet at 4550\AA, 
the T$_{\rm eff}$ and $\log g$ determined through the  H$\beta$ line.} 
\label{vr} 
\centering                
\begin{tabular}{c r c c }     
\hline                  
\hline            
HJD         & V$_{\rm rad}$~~~~~ & T$_{\rm eff}$ & $\log g$ \\
(2554041.0+) & (km s$^{-1}$)~~ &     (K)       &                 \\
\hline                                                                                     
0.2510 & $-$19.70\,$\pm$\,0.97 & 24500\,$\pm$\,310 & 3.95\,$\pm$\,0.04  \\
0.2587 & $-$17.86\,$\pm$\,0.85 & 24150\,$\pm$\,240 & 3.87\,$\pm$\,0.05  \\
0.2674 & $-$16.68\,$\pm$\,0.40 & 24300\,$\pm$\,240 & 3.89\,$\pm$\,0.05  \\
0.2746 & $-$13.67\,$\pm$\,0.49 & 24250\,$\pm$\,220 & 3.90\,$\pm$\,0.05  \\
0.2819 & $-$10.01\,$\pm$\,0.35 & 24150\,$\pm$\,220 & 3.88\,$\pm$\,0.05  \\
0.2892 &  $-$6.86\,$\pm$\,0.28 & 24050\,$\pm$\,210 & 3.88\,$\pm$\,0.05  \\
0.2964 &  $-$3.95\,$\pm$\,0.40 & 23800\,$\pm$\,200 & 3.81\,$\pm$\,0.05  \\
0.3040 &     0.76\,$\pm$\,0.19 & 23700\,$\pm$\,160 & 3.78\,$\pm$\,0.04  \\
0.3113 &     4.89\,$\pm$\,0.30 & 23750\,$\pm$\,170 & 3.78\,$\pm$\,0.04  \\
0.3186 &     8.18\,$\pm$\,0.82 & 23700\,$\pm$\,160 & 3.77\,$\pm$\,0.04  \\
0.3261 &    10.63\,$\pm$\,0.55 & 23850\,$\pm$\,210 & 3.81\,$\pm$\,0.05  \\
0.3334 &    11.97\,$\pm$\,0.64 & 23850\,$\pm$\,210 & 3.85\,$\pm$\,0.06  \\
0.3406 &    12.18\,$\pm$\,0.26 & 23850\,$\pm$\,220 & 3.84\,$\pm$\,0.06  \\
0.3515 &    10.70\,$\pm$\,0.22 & 24150\,$\pm$\,210 & 3.88\,$\pm$\,0.05  \\
0.3588 &     8.24\,$\pm$\,0.50 & 24400\,$\pm$\,220 & 3.92\,$\pm$\,0.04  \\
0.3660 &     4.95\,$\pm$\,0.28 & 24200\,$\pm$\,210 & 3.89\,$\pm$\,0.05  \\
0.3733 &     1.75\,$\pm$\,1.01 & 24300\,$\pm$\,240 & 3.90\,$\pm$\,0.05  \\
0.3806 &  $-$1.00\,$\pm$\,0.99 & 24400\,$\pm$\,260 & 3.92\,$\pm$\,0.04  \\
0.3878 &  $-$5.13\,$\pm$\,1.42 & 24250\,$\pm$\,250 & 3.89\,$\pm$\,0.06  \\
0.3955 &  $-$8.74\,$\pm$\,2.13 & 24200\,$\pm$\,250 & 3.86\,$\pm$\,0.06  \\
0.4028 & $-$12.14\,$\pm$\,1.04 & 24200\,$\pm$\,250 & 3.89\,$\pm$\,0.06  \\
0.4100 & $-$15.52\,$\pm$\,0.71 & 24350\,$\pm$\,250 & 3.91\,$\pm$\,0.06  \\
0.4173 & $-$18.43\,$\pm$\,0.29 & 24350\,$\pm$\,240 & 3.90\,$\pm$\,0.05  \\
0.4246 & $-$20.11\,$\pm$\,0.94 & 24350\,$\pm$\,210 & 3.92\,$\pm$\,0.04  \\
0.4318 & $-$20.13\,$\pm$\,0.88 & 24300\,$\pm$\,210 & 3.90\,$\pm$\,0.05  \\
0.4475 & $-$19.34\,$\pm$\,1.26 & 24050\,$\pm$\,210 & 3.87\,$\pm$\,0.05  \\
0.4548 & $-$15.45\,$\pm$\,2.05 & 24050\,$\pm$\,230 & 3.87\,$\pm$\,0.05  \\
0.4620 & $-$14.02\,$\pm$\,1.02 & 23850\,$\pm$\,210 & 3.83\,$\pm$\,0.05  \\ 
0.4693 & $-$11.86\,$\pm$\,1.57 & 23900\,$\pm$\,210 & 3.84\,$\pm$\,0.05  \\
0.4766 &  $-$8.07\,$\pm$\,0.89 & 23800\,$\pm$\,190 & 3.81\,$\pm$\,0.05  \\
0.4838 &  $-$4.53\,$\pm$\,0.92 & 23800\,$\pm$\,190 & 3.81\,$\pm$\,0.05  \\
\hline  
\end{tabular}
\end{table}

\section{Observations and data reduction}
\label{observ}
Spectroscopic observations of $\beta$ Cep have been carried out at the 91~cm 
telescope of the {\it INAF - Osservatorio Astrofisico di Catania} at 
R\,=\,20\,000 in the range 4300-6800 {\AA} \citep{catanzaro08}. 

Spectra were acquired during the night of November, 
the 1$^{st}$ 2006, starting at JD\,=\,2454041.2510, for a total of 31 exposures. 
Taking into account the pulsational period P\,=\,0.19048678 days \citep{chap85},
we set the exposure time to 10 minutes, $\approx$~4$\%$ of the pulsational period,
to avoid phase smearing effects. 
The stellar spectra, calibrated in wavelength and with the continuum normalised
to a unity level, were obtained using standard data reduction procedures for 
spectroscopic observations within the NOAO/IRAF package \citep{catanzaro08}.
The signal-to-noise ratio of our spectra always resulted above 200. 

Instrumental systematic errors on radial velocities have been corrected
observing stars with 
constant and well-known radial velocity taken from the list of standard stars 
published by \citet{udry99}: HD\,186791 and HD\,206778.

\section{Radial velocity variability}
\label{vrad}
Radial velocities (V$_{\rm rad}$) of $\beta$ Cep, as it is common
for this class of pulsating stars, have been measured by means of the 
Si{\sc iii} triplet at 4552, 4567 and 4574\,{\AA}. These are strong
lines almost insensitive to temperature variations \citep{deridder02}
and presenting no significative blending and. 

\begin{figure}
   \centering
   \includegraphics[width=9cm]{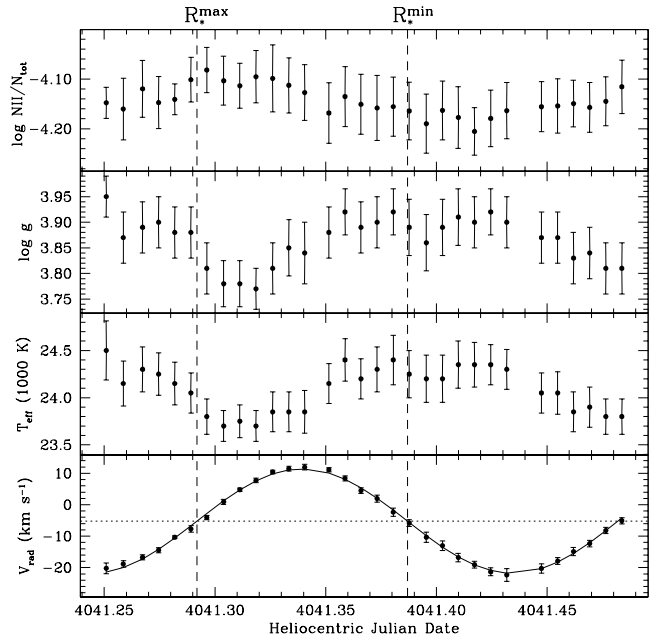}
   \caption{From bottom to top, the radial velocity measured from Si{\sc iii} triplet, effective 
            temperature and gravity obtained by fitting H$\beta$ line profiles,
            and nitrogen abundance are plotted as a function of time.
            Vertical dashed lines represent the instants of maximum and minimum
            radius. The over-imposed sinusoidal fit of velocities has been 
            calculated adopting pulsational period derived by \citet{chap85}.}
   \label{teff_var}
   \end{figure}

Errors on V$_{\rm rad}$ have been computed assuming that the main source of
error is the statistical noise in the observed spectrum. 
Velocities are reported in Tab.~\ref{vr} together with the 
respective Heliocentric Julian Date and plotted in the bottom panel of
Fig.\,\ref{teff_var}. The parameters of the over-imposed sinusoidal fit have been 
computed adopting the pulsational period by \citet{chap85}.
The results of fitting procedure were: an average value
$\gamma_0$\,=\,$-$4.8\,$\pm$\,0.1~km~s$^{-1}$ and amplitude
K\,=\,16.3\,$\pm$\,0.2~km~s$^{-1}$.

\section{Average stellar parameters}
\label{atmos}
The first step to highlight the variability of effective temperature
(T$_{\rm eff}$), surface gravity ($\log g$) and microturbulence
($\xi$) in $\beta$\,Cep is the ascertainment of their average
behaviour. 

\citet{Leone97} have demonstrated that the effective
temperature and gravity inferred from Balmer lines have to be determined
simultaneously with abundances. These 
define the metal opacity and with it the dependence of the temperature
on the optical depth \citep{Leone96}. Moreover, the abundances
define the Stark broadening of Balmer lines through the electron pressure.
Similarly, the effective temperature and gravity of a star have to be
determined simultaneously with the microturbulence. This velocity field
defines the metal opacity through the line-blocking.
This is an effect particularly important for a hot star
whose flux is peaked in the ultraviolet where the largest concentration
of atomic transitions appear. As to $\beta$\,Cep,
\citet{morel06} determined a rather high (6$\pm$3 km s$^{-1}$)
microturbulence certainly able to largely desaturate metal lines.

To establish the atmosphere parameters of $\beta$ Cep,
we implemented the iterative procedure presented in \citet{Leone97}
to include the microturbulence. In particular we compare the observed and 
theoretical profiles of H$\gamma$ and H$\beta$ lines by minimising the 
residuals. Abundances are derived froma comparison between the observed 
average spectrum and SYNTHE spectra.
Following \citet{Erspamer03}, the observed spectrum
has been divided to intervals of about 100 \AA\, and abundances are given
by the average values in different intervals. Errors represent the
standard deviations. The microturbulence velocity is determined
demanding that the abundances derived from 14 N {\sc ii} and 17 O {\sc ii}
unblended lines are independent of equivalent widths. In this case, 
the abundances are derived by means of the WIDTH9 code \citep{kur81}.

As regards the effective temperature and gravity we confirm the values found by
\citet{catanzaro08} (see Fig.~1 of that paper): T$_{\rm eff}$\,=\,24000\,$\pm$\,250~K 
and $\log g$\,=\,3.91\,$\pm$\,0.05, while $\xi$\,=\,8.1\,$\pm$\,0.9 km s$^{-1}$
 and the abundances listed in Table\,\ref{atmos_sum}.

As a general result, we derive abundances close to the solar values 
\citep{asplund05}, with the exception of carbon and iron for
which a slight under-abundance has been inferred. The slight under-abundance 
of magnesium, here derived using only one line, has to be confirmed.

\begin{table}
\begin{minipage}[t]{\columnwidth}
\caption{We report the atmosphere parameters of $\beta$ Cep here determined: 
T$_{\rm eff}$, $\log g$, microturbulence ($\xi$) and abundances.
$N$ is the number of identified lines for any element.
For comparison, we report the values by: \citet{morel06} (M06), 
\citet{heyn94} (H94), \citet{niec05} (N05) and the solar abundances by 
\citet{asplund05} (A05).} 
\label{atmos_sum} 
\centering                 
\renewcommand{\footnoterule}{}  
\begin{tabular}{l ccc}     
\hline                  
\hline            
 & {\it This study}       & \multicolumn{1}{c}{\it M06} & {\it ~H94}\hspace{0.7cm}{\it N05}~~~~~~\\
\hline       
T$_{\rm eff}$ (K)   &24000$\pm$250  & 26000$\pm$1000 & 24550\hspace{0.32cm}  24150$\pm$350\\
$\log g$~~~         &3.91$\pm$0.05 &  3.70$\pm$0.15  & 3.772~~~\hspace{0.5cm}  3.69~~~~~~ \\
$\xi$\,(km s$^{-1}$) &8.1$\pm$0.9  &     6$\pm$3    &               \\
\hline                  
\hline
El~~~~ N& \multicolumn{3}{c}{$\log N_{\rm El}/N_{\rm tot}$~~~~~~~~~~~~~~~~~~~~}  \\
\multicolumn{2}{l}{\it ~~~~~~~~~~~~~This study} &   \multicolumn{1}{c}{\it M06} & {\it A05} \\
\hline
\multicolumn{2}{l}{C~~~~~8\hspace{0.35cm}$-$3.92\,$\pm$\,0.05}   & $-$4.01\,$\pm$\,0.10 & $-$3.64\,$\pm$\,0.05\\
\multicolumn{2}{l}{N~~~~31\hspace{0.32cm}$-$4.20\,$\pm$\,0.07}   & $-$4.12\,$\pm$\,0.13 & $-$4.25\,$\pm$\,0.06\\
\multicolumn{2}{l}{O~~~~41\hspace{0.32cm}$-$3.39\,$\pm$\,0.07}   & $-$3.56\,$\pm$\,0.14 & $-$3.37\,$\pm$\,0.05\\
\multicolumn{2}{l}{Ne~~~~2\hspace{0.35cm}$-$4.00\,$\pm$\,0.13}   &                      & $-$4.19\,$\pm$\,0.06\\
\multicolumn{2}{l}{Mg~~~1\footnote{Mg{\sc ii} $\lambda$ 4481 {\AA}}~~$-$4.85\,$\pm$\,0.10} &  $-$4.72\,$\pm$\,0.21 & $-$4.50\,$\pm$\,0.09\\
\multicolumn{2}{l}{Al~~~~5~~~~$-$5.80\,$\pm$\,0.30}   & $-$6.02\,$\pm$\,0.16 & $-$5.66\,$\pm$\,0.06\\
\multicolumn{2}{l}{Si~~~10~~~~$-$4.70\,$\pm$\,0.19}   & $-$4.92\,$\pm$\,0.23 & $-$4.52\,$\pm$\,0.04\\
\multicolumn{2}{l}{S~~~~~3~~~~$-$4.90\,$\pm$\,0.20}   & $-$4.89\,$\pm$\,0.37 & $-$4.89\,$\pm$\,0.05\\
\multicolumn{2}{l}{Fe~~14\hspace{0.43cm}$-$4.81\,$\pm$\,0.13}   & $-$4.79\,$\pm$\,0.23 & $-$4.58\,$\pm$\,0.05\\
\hline 
\end{tabular}
\end{minipage}
\end{table}

\section{Variability of atmosphere properties}
The spectral synthesis of the average spectrum of $\beta$ Cep
shows the presence of strong oxygen lines in the wings of the
H$\gamma$ line. The equivalent width variability of oxygen lines
 with the pulsation period (Fig.\,\ref{var_ew})
would complicate and false the result of
any attempt to ascertain the variability of T$_{\rm eff}$ and $\log g$
on the basis of this line. For this reason, we prefer to
base this study on the H$\beta$ line only.

\begin{figure}
   \centering
   \includegraphics[height=15cm]{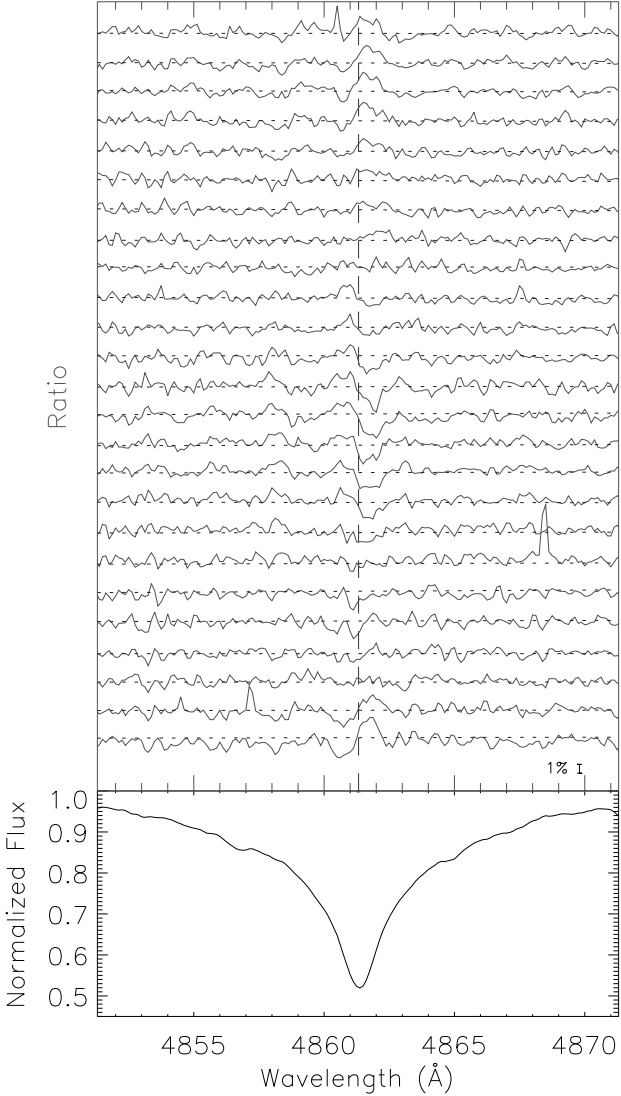}
   \caption{The H$\beta$ profile showed in the bottom is the
           average of single profiles acquired through a complete 
           oscillation period. The upper panel
           shows the ratios among single profiles and the average one.  The
           1~$\%$ error bar is also reported.}
   \label{rapp_hb}
   \end{figure}

Before determining the effective temperature and gravity
of any single spectrum, we check evidence for a variability
in the H$\beta$ profile  
in addition to the radial velocity one: each spectrum has
been Doppler corrected for the velocity determined by means of
silicon lines and divided by the average H$\beta$ profile.
These ratios are shown in the upper panel of Fig.\,\ref{rapp_hb}.
The usual spectral distortion due to pulsations characterises only
the line core; this is here interpreted as evidence for a
larger velocity presented by the outer layers.
The variability of the radial velocity determined through the core of
the H$\beta$ line presents an amplitude of 18.7$\pm$0.3 km\,s$^{-1}$ and an
average value of $-3.3\pm$0.2 km\,s$^{-1}$. 

\subsection{Microturbulence}
To point out any possible variability of stellar parameters with the
pulsational period, we applied the previous iterative procedure to any single
spectrum of $\beta$ Cep.

If we consider the effective temperature, gravity
and microturbulence as free parameters, the microturbulence from
nitrogen lines along the pulsation cycle results variable between 5.4 and 8.4 km s$^{-1}$.
Differently, the oxygen lines give a constant value equal to
8.1$\pm$0.9 km s$^{-1}$. 

Source functions of the considered 14 nitrogen and 17 oxygen lines are
equally distributed in optical depth
between $\log\tau_{5000 \AA} = -0.85$ and $-$2.06. These rule out any changing
of the microturbulence with the atmospheric height.

  \begin{figure}
   \centering
   \includegraphics[width=9cm]{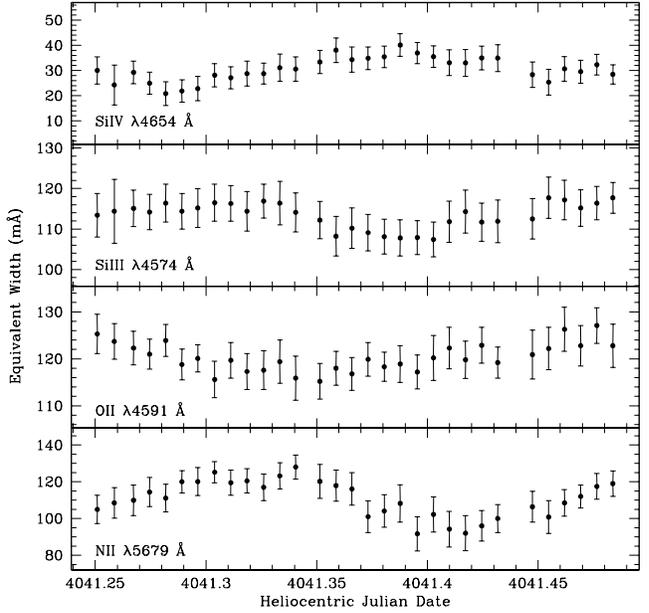}
   \caption{Equivalent width of representative oxygen, nitrogen and silicon
    lines are reported to show their variability with the pulsation period.
    It is worth noting that O {\sc ii} changes out of phase compared to
    N {\sc ii} and Si {\sc iii} lines, while it is in phase with Si {\sc iv}.}
   \label{var_ew}
   \end{figure}

\begin{table}
\caption{Theoretical equivalent widths in m\AA\, for three Si {\sc iii} lines,
computed by means of WIDTH9, for three values of the microturbulence.
Computations are based on the previous atmosphere parameters of $\beta$ Cep.} 
\label{equiv} 
\centering                
\begin{tabular}{c|rrl}     
\hline                  
\hline            
\multicolumn{1}{r}{Wavelength (\AA)}& $\xi$ = 5.4 & 6.9 & ~8.4 (km\,s$^{-1}$)\\
\hline       
4552 & 160 & 183 & 206 \\
4567 & 141 & 161 & 180 \\
4574 & 104 & 118 & 131 \\
\hline      
\end{tabular}
\end{table}

The number and equivalent widths of the silicon lines
in our spectra are not enough to measure the microturbulence. However, the
Si{\sc iii} triplet at 4552, 4567 and 4574\,{\AA} is very sensitive to
this velocity field and rules out a microturbulence range larger than 
1.0 km s$^{-1}$. Table\,\ref{equiv} shows that for the microturbulence
variation deduced from nitrogen lines, the silicon triplet would present
a 25\% variability of equivalent widths much larger than the 7\% observed
(Fig.\,\ref{var_ew}). We conclude that the microturbulence variability determined
from nitrogen lines is a local minimum of the previous $\chi^2$
function and we assume a constant value for this velocity equal
to the one we obtained from the average spectrum: $\xi$ = 8.1
km s$^{-1}$.

\subsection{T$_{\rm eff}$ and $\log g$ variability}
We find that the effective temperature and gravity of $\beta$\,Cep are variable 
with the pulsation period (Tab.~\ref{vr}).
Fig.\,\ref{teff_var} shows that T$_{\rm eff}$ changes, almost sinusoidally
from $\sim$23700~K to 24400~K, in phase with the $\log g$ variability
from 3.76 to 3.93. 

For a constant $\xi$, the variability of equivalent widths with the 
pulsation period is also indicative of the temperature behaviour along a
cycle. Fig.\,\ref{var_ew} shows that N {\sc ii} and Si {\sc iii} lines are 
in phase and that they change out of phase compared to O {\sc ii} and Si {\sc iv}. 
We have computed the expected equivalent widths of the representative lines
of Fig.\,\ref{var_ew} for atmosphere models with  20000~$\le T_{\rm eff} \le$~30000 K,
$\log g$ = 3.91, $\xi$ = 8.1 km s$^{-1}$ and abundance listed in Table 2.
It appears (Fig.\,\ref{teff_var}) that the observed behaviour of these lines
is due to a variation of the effective temperature between 25000 and 26000 K.
That is a result in agreement with \citet{morel06} who derived an
effective temperature of 26000 K from the ratio of Si {\sc iv} and Si {\sc iii}
line strength. The top panel of Fig.~\ref{teff_var} shows the N {\sc ii} abundance 
computed along the pulsational phase, considering the corresponding values
of T$_{\rm eff}$ and $\log g$ and keeping constant the value of $\xi$ 
(8.1 km s$^{-1}$). This reflects the variations of the ionisation
state of nitrogen.

\section{Conclusions}
With the aim of determining the variations of the effective temperature,
surface gravity and microturbulence velocity that occur in hot pulsating
stars during the pulsational cycle, we performed time resolved R\,=\,20000
spectroscopic observations of the prototype $\beta$~Cep along a complete
period coverage. 

As a first step, we confirm the T$_{\rm eff}$ and $\log g$ found in
a previous paper by \citet{catanzaro08} and we measure $\xi$\,=\,8.1\,$\pm$\,0.9~km~s$^{-1}$,
from a sample of single lines of O{\sc ii}. With these values,
the abundances resulted to be solar for nitrogen, oxygen, neon,
aluminium, silicon and sulfur, while slight under-abundances have
been inferred for carbon, magnesium and iron.

We find that along a pulsation cycle, after a correction for
the pulsational Doppler shift, the H$\beta$ line profile changes
are matched assuming a variability of the effective temperature
in the range of 23700 to 24400 K and a variability of gravity
in the range of 3.76 to 3.93. The amplitude of $\approx$~0.08~dex
measured in the gravity variation could lead the reader to the 
erroneous belief that it corresponds to a radius variation larger 
than the one computed by \citet{aerts94}, via cinematic considerations. 
However, the variability of $\log g$ here determined for $\beta$~Cep
is not a consequence of changes in the surface gravity because of expansions
and compressions of the star as a whole. We ascribed the $\log g$ 
variability to a change of the electron pressure \citep{Gray92}
induced by temperature, that modulates the ionisation state of metals.
Thus, it should not be linked directly to the radius of the star.

Metal lines are characterised by equivalent widths variable
with the pulsation cycle. These are explained by a variable effective 
temperature between 25000 and 26000 K as a consequence of the changed
ionisation state. It could be that the most extended formation region
of the H$\beta$ line is at the origin of the lower temperature, in agreement 
with the 15\% larger amplitude presented by the radial velocity measured 
through the H$\beta$ core.

\begin{figure}
   \centering
   \includegraphics[width=8cm]{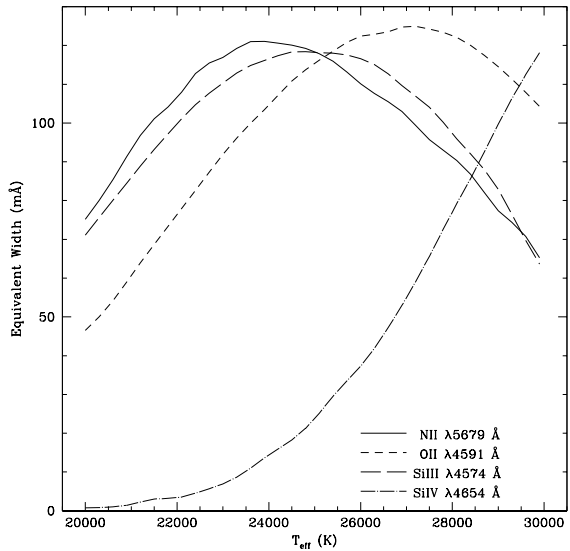}
   \caption{Equivalent widths as a function of effective temperature.}
   \label{ew_T}
   \end{figure}

As to the possibility that the microturbulence depends on the
compression and expansion of the atmospheric layers, we find that
it is constant within a 1.0 km~s$^{-1}$.

We conclude that pulsations modify the dependence of the temperature 
and pressure on the optical depth in the atmosphere of $\beta$~Cep. This 
mimics a variability of the effective temperature and gravity when a small 
range of wavelengths is selected.  We hope that this information can help 
in modeling the variability of spectral lines that is not only due to the 
pulsation velocity field \citep{dupret02}. 

Along the expansion phase, the negative radial velocity is expected
to increase up to a null value just in coincidence with
the maximum of the stellar radius. At this phase of minimum
compression, the photosphere is expected at the temperature minimum.
Differently, we observe a delay in the temperature minimum about one
tenth of the pulsation period. According to \citet{deridder02},
this extra-phase shift is due to the non-adiabatic condition that occurs
in the  pulsating layers. 

\section*{Acknowledgments}
The authors wish to thank Luigia Santagati for the english revision of the
text.

\label{lastpage}

\end{document}